\documentclass[10pt,twocolumn,oneside,journal]{IEEEtran}
\usepackage{cite}
\usepackage{graphicx}
\usepackage{amsmath}
\usepackage{times}
\usepackage{latexsym}
\usepackage{graphicx}
\usepackage{epstopdf}
\usepackage{bm}
\usepackage{amssymb}
\usepackage[center]{caption2}
\usepackage{stfloats}
\usepackage{cases}
\usepackage{array}
\usepackage{setspace}
\usepackage{fancyhdr}
\usepackage{fancyhdr}
\usepackage{algorithm}
\usepackage{algpseudocode}
\usepackage{textcomp}
\usepackage{wasysym}
\usepackage{url}
\usepackage{xcolor}
\usepackage{verbatim}
\usepackage{stfloats}
\usepackage{float}

\allowdisplaybreaks[4]

\newcaptionstyle{mystyle1}{%
  \centering TABLE  \captiontext \par}
\captionstyle{mystyle1}
\newcaptionstyle{mystyle2}{%
  \captionlabel $.$ \: \captiontext \par}
\captionstyle{mystyle2}
\newcaptionstyle{mystyle3}{%
  \captionlabel $.$ \: \captiontext \par}
\captionstyle{mystyle3}

\renewcommand{\QED}{\QEDopen}

\begin{document}

\title{Distributed Edge Caching in Ultra-dense Fog Radio Access Networks: A Mean Field Approach}


\author{Yabai~Hu,
Yanxiang~Jiang,~\IEEEmembership{Member,~IEEE},
Mehdi~Bennis,~\IEEEmembership{Senior~Member,~IEEE}, and Fu-Chun~Zheng,~\IEEEmembership{Senior~Member,~IEEE},
\thanks{Y. Hu and Y. Jiang are with the National Mobile Communications Research Laboratory, Southeast University, Nanjing 210096, China,
the State Key Laboratory of Integrated Services Networks, Xidian University, Xi'an 710071, China,
and the Key Laboratory of Wireless Sensor Network $\&$ Communication, Shanghai Institute of Microsystem and Information Technology, 
Chinese Academy of Sciences, Shanghai 200050, China. (e-mail: yxjiang@seu.edu.cn)}
\thanks{M. Bennis is with the Centre for Wireless Communications, University of Oulu, Oulu 90014, Finland. (e-mail: bennis@ee.oulu.fi)}
\thanks{F. Zheng is with the School of Electronic and Information Engineering, Harbin Institute of Technology, Shenzhen 518055, China, and the National Mobile Communications Research Laboratory, Southeast University, Nanjing 210096, China. (e-mail: fzheng@ieee.org)}
}
\maketitle

\begin{abstract}
In this paper, the edge caching problem in ultra-dense fog radio access networks (F-RAN) is investigated. Taking into account time-variant user requests and ultra-dense deployment of fog access points (F-APs), we propose a dynamic distributed edge caching scheme to jointly minimize the request service delay and fronthaul traffic load. Considering the interactive relationship among F-APs, we model the caching optimization problem as a stochastic differential game (SDG) which captures the temporal dynamics of F-AP states and incorporates user requests status. The SDG is further approximated as a mean field game (MFG) by exploiting the ultra-dense property of F-RAN. 
In the MFG, each F-AP can optimize its caching policy independently through iteratively solving the corresponding partial differential equations without any information exchange with other F-APs. The simulation results show that the proposed edge caching scheme outperforms the baseline schemes under both static and time-variant user requests.
\end{abstract}

\begin{keywords}
Fog radio access networks, distributed edge caching, mean field game, request service delay, fronthaul traffic load.
\end{keywords}

\section{Introduction}
The increasing number of mobile devices and surging user demand for multi-media service have  triggered massive amounts of data in wireless networks, which brings about huge pressure on fronthaul links. Fog radio access networks (F-RAN) can address this problem by allocating caching and computing resources to fog access points (F-APs) deployed ultra-densely at the network edge. In ultra-dense F-RAN, large deployment density enables F-APs to provide large coverage area and high quality wireless links, while edge caching alleviates fronthaul traffic pressure and decreases request service delay by bringing contents closer to users \cite{Fog2, Dense, Cache}. 

Due to the limited caching space of F-APs, the optimization of caching placement is a major issue in F-RAN. The caching placement design can be centralized or distributed. In centralized schemes, caching policies are generally constructed by a control center that collects information of the entire network for seeking a global optimal solution. However, the difficulty in information acquirement and global optimization renders centralized schemes impractical in ultra-dense networks. In distributed caching schemes, on the contrary, edge caching nodes can independently optimize their caching policies based solely on local information. In \cite{Distr1}, a belief propagation (BP)-based distributed caching algorithm was put forward, where each base station computes the caching placement matrix through message passing between nodes. In \cite{Distr3}, the cost of fronthaul usage was minimized by using a distributed algorithm based on alternating direction method of multipliers (ADMM). The authors of \cite{Distr4} modeled the network dynamics in a game-theoretic approach and proposed a distributed caching scheme to minimize the backhaul traffic load. A Markov-chain-based optimization framework was established to minimize the macro-cell load by considering high-mobility scenarios in \cite{Distr5}.

However, all the above works are based on the assumption of a static content popularity distribution, whereas the user requests for popular contents change over time in practice. Hence, a dynamic distributed edge caching scheme to meet the time-variant demand of users is needed. In \cite{Dyn1} and \cite{Dyn2}, the dynamic evolution of content popularity was investigated using differential equations and Markov chains, respectively. 
Nevertheless, none of these works account for both request service delay and fronthaul traffic load.

Motivated by the aforementioned discussions, we propose a distributed edge caching scheme adaptive to dynamic user requests in ultra-dense F-RAN. To jointly minimize the request service delay and fronthaul traffic load, a stochastic differential game (SDG) is formulated to model the edge caching problem. The SDG captures the interactions among F-APs and the dynamics of F-AP states, incorporating the status of temporal user requests. Due to the high computational complexity of obtaining the solution, the SDG is approximated as a mean field game (MFG) to solve this challenging problem by utilizing the ultra-dense property of F-RAN. In the MFG, the individual states of all the F-APs can be approximated by a statistical mean field distribution. Then, each F-AP is able to determine the dynamic optimal caching policy based on its local state and the mean field distribution without extra information exchange with the cloud server or adjacent F-APs. 


\section{System Model}

\subsection{Network Model}

    Consider an edge caching enabled F-RAN consisting of $I$ F-APs, $K$ users, and one remote cloud server, in which the F-APs are ultra-densely deployed and connected with the cloud server via fronthaul links. The F-AP set and the user set are denoted by $\mathcal{\boldsymbol{I}} = \left\{ {1,2, \cdots ,{i}, \cdots ,I} \right\}$ and $\mathcal{\boldsymbol{K}} = \left\{ {1,2, \cdots ,{k}, \cdots ,K} \right\}$, respectively. Assume that F-APs are uniformly distributed, users are randomly distributed, and each user associates with a default serving F-AP which is the geographically nearest one. {The users with the same default serving F-AP ${i}$ form the serving user cluster ${{\bm{U}}^i}$, and each F-AP serves at most one user at a time.} Assume that users request files from the content library $\mathcal{\boldsymbol{F}} = \left\{ {{f_1}, {f_2}, \cdots ,{f_n}, \cdots ,{f_N}} \right\}$. Without loss of generality, each file is assumed to have the same size of $S$ bits. The request arrival in consecutive time slots is modeled as a Poisson process, and the $t$th time slot is denoted as time $t$ for the sake of simplicity. Users receive the requested files from the corresponding F-APs through wireless links, and the transmission rate from F-AP ${i}$ to its serving user ${k}$ at time $t$ can be expressed as follows:
    \begin{equation}
      {R_{i,k}}\left( t \right) = W{\log _2}\left( {1 + \frac{{\left| {{h_{i,k}}\left( t \right)} \right|^2}{P_i}}{{{\sigma ^2} + \sum\nolimits_{j \ne i,j \in \mathcal{\boldsymbol{I}}} {{\left| {{h_{j,k}}\left( t \right)} \right|^2}{P_j}} }}} \right),
    \end{equation}
    where $W$ is the transmission bandwidth, ${P_i}$ is the transmission power of F-AP ${i}$, ${\sigma ^2}$ is the noise power, {and $h_{i,k}\left( t \right)$ is the channel coefficient between F-AP ${i}$ and user ${k}$ at time $t$. By considering a time-varying fading channel, the evolution of ${h_{i,k}}\left( t \right)$ can be characterized by a mean-reverting Ornstein-Uhlenbeck process as follows \cite{Channel-Chapter}:}
    \begin{equation}\label{2-1}
      {\rm{d}}{h_{i,k}}\left( t \right) = \frac{1}{2}\alpha \left( {{\mu _h} - {h_{i,k}}\left( t \right)} \right){\rm{d}}t + {\sigma _h}{\rm{d}}{B_{i,k}}\left( t \right),
    \end{equation}
    {where ${\mu _h}$ and \(\sigma _h\) with positive values characterize the long-term mean and variance of the process, respectively,} $\alpha$ is the speed of the process towards the long-term mean, and ${ B_{i,k}}\left( t \right)$ is a standard Brownian motion that depicts the stochastic dynamics of wireless channels.

\subsection{Content Caching and Delivery Model}

Regarding the content processing policy of each F-AP, each request time slot is divided into two phases: caching placement phase and content delivery phase. The storage space of the cloud server is assumed to be considerably large that contains the whole content library, while each F-AP is equipped with limited caching space that can store up to $C$ bits.

In the caching placement phase, F-AP $i$ fetches certain fractions of files through the fronthaul link according to its caching policy which is denoted by the caching vector ${{\bm{c}}_i}\left( t \right)$ of length $N$. The vector element ${c_{i,n}}\left( t \right) \in \left[ {0,1} \right]$ denotes the instantaneous caching rate of file ${f_n}$, representing the proportion of the file to be cached by downloading through the fronthaul link at time $t$. The cache state of F-AP ${i}$ with respect to (w.r.t.) ${f_n}$ is denoted by ${s_{i, n}}\left( t \right) \in \left[ {0,S} \right]$, which is the instantaneous number of bits stored in its caching space at time $t$. {Assume that file $f_n$ will be discarded by the F-AP  when it is rarely requested, and the less requests for file ${f_n}$, the faster the file to be discarded from the caching space.} The dynamics of the cache state can then be expressed as follows:
 \begin{equation}\label{2-2}
   {\rm{d}}{s_{i,n}}\left( t \right) = {S\left( {{c_{i,n}}\left( t \right) - {e^{a - 1}}{a^{{q_{i,n}}\left( t \right)}}} \right)}{\rm{d}}t,
\end{equation}
where ${q_{i,n}}\left( t \right)$ denotes the number of requests for ${f_n}$ from the serving user cluster of F-AP ${i}$ at time $t$. {Note that the incrementing rate of the cache state is represented by the caching rate ${c_{i,n}}\left( t \right)$, while the decrementing rate is represented by  ${{e^{a - 1}}{a^{{q_{i,n}}\left( t \right)}}}$, which is a descending function of ${q_{i,n}}\left( t \right)$ with the parameter $a \in \left( {0,1} \right)$ that controls the steepness of the function. Therefore, a small request number for $f_n$ results in a large discarding rate, which adapts to real-time requests.} 

{In the content delivery phase, each F-AP processes the user requests during the current time slot according to the cache states of its own and the other F-APs, which can be classified into three cases. For case 1,} a user that requests for file ${f_n}$ associates with its default serving F-AP if the file is already cached.
{Otherwise, for case 2,} the user seeks other adjacent F-APs that have cached ${f_n}$ and associates with the one that provides the largest transmission rate \cite{Distr1}. {For case $3$, when none of the F-APs has cached the requested file,} the user will then still associate with the default serving F-AP.
{Considering that the cache state ${s_{i,n}}\left( t \right)$ is a continuous variable ranging from 0 to $S$, here we assume that file $f_n$ is thought to be cached by the default serving F-AP as long as ${s_{i,n}}\left( t \right) \ge \frac{1}{2}S$, while file $f_n$ is thought to be cached by the other F-APs only when ${s_{ - i,n}}\left( t \right) = S$, where ${s_{ - i,n}}\left( t \right)$ denotes the cache state of the other possible serving F-APs.}

Given the system model introduced above, the overall state of F-AP $i$ can be represented by its channel state in (\ref{2-1}) and cache state in (\ref{2-2}). Let ${{\bm{h}}_i} = {\left[ {{h_{i,1}},{h_{i,2}}, \cdots ,{h_{i,K}}} \right]^{\rm{T}}}$ and ${{\bm{s}}_i} = {\left[ {{s_{i,1}},{s_{i,2}}, \cdots ,{s_{i,N}}} \right]^{\rm{T}}}$  denote the channel state and cache state vector of F-AP $i$, respectively. Then, the state vector of F-AP $i$ can be denoted by ${{\bm{x}}_{i}} = \left( {{{\bm{h}}_{i}},{\bm{s}_{i}}} \right)$.

The objective of this paper is to find the optimal edge caching policy to jointly minimize the request service delay and fronthaul traffic load for each F-AP.

\section{The Proposed Dynamic Distributed Edge Caching Scheme}
\newcounter{TempEqCnt} 
\setcounter{TempEqCnt}{\value{equation}} 
\setcounter{equation}{3} 
\begin{figure*}[t]
\begin{equation}\label{Delay}
\begin{split}
{D_{i,n}}\left( t \right) = \sum_{\substack{k \in {{\bm{U}}^i}\\
\scriptstyle {{\rm req}_k}\left( t \right) = n}} &\left(
\left( {{C_1}\left( {t,{s_{i,n}}\left( t \right)} \right)} + {C_3}\left( {t,{s_{i,n}}\left( t \right),{s_{ - i,n}}\left( t \right)} \right) \right)\left( {\frac{S}{{{R_{i,k}}\left( t \right)}} + \frac{{S - {s_{i,n}}\left( t \right)}}{{{R^F}}}} \right)\right.\\
 &\left.+ {C_2}\left( {t,{s_{i,n}}\left( t \right),{s_{ - i,n}}\left( t \right)} \right)\frac{S}{{{R_{ - i,k}}\left( t \right)}} \right).
 \end{split}
\end{equation}
\begin{equation}\label{Load}
  \begin{split}
{O_{i,n}}\left( t \right) &= \left( {{\eta _1}{c_{i,n}}\left( t \right) + {\frac{1}{2}{\eta _2}}{c^2_{i,n}}\left( t \right)} \right) + \eta {q_{i,n}}\left( t \right) \left( {S - {s_{i,n}}\left( t \right)} \right)\left( {{C_1}\left( {t,{s_{i,n}}\left( t \right)} \right) + {C_3}\left( {t,{s_{i,n}}\left( t \right),{s_{ - i,n}}\left( t \right)} \right)} \right).
\end{split}
\end{equation}
\hrulefill
\end{figure*}
\setcounter{equation}{\value{TempEqCnt}} 
 In this section, we firstly define the cost function for each F-AP. Then, we formulate the stochastic differential game to model the caching optimization problem. Finally, a distributed mean field edge caching scheme that can adapt to time-variant user requests is proposed.

\subsection{Cost Function}

Let \({J_{i,n}}\left( t \right)\) denote the total cost of F-AP ${i}$ for file ${f_n}$ at time $t$, which is composed of request service delay and fronthaul traffic load.

\subsubsection{Request Service Delay}
 Let ${D_{i,n}}\left( t \right)$ denote the delay cost of F-AP ${i}$ with respect to file ${f_n}$, which is the total service delay of the requests for ${f_n}$ from its serving user cluster ${{\bm{U}}^i}$ during the $t$th time slot. Note that the request service delay refers to the total elapsed time from the file being requested to the user being successfully served, which is composed of wireless transmission delay and possible fronthaul retrieving delay. 

  The request service delay is expressed in (\ref{Delay}) at the top of this page, where ${{\rm req}_k}\left( t \right)$ denotes the requested file of user $k$ at time $t$, {${R^F}$ denotes the fronthaul transmission rate, and ${{{R}_{ - i,k}}\left( t \right)}$ denotes transmission rate provided by the other serving F-AP in case 2. The occurrence conditions for the three cases introduced in Section \uppercase\expandafter{\romannumeral2}-B can be expressed as follows:
   \begin{equation*}
   \begin{split}
   &{C_1}\left( {t,{s_{i,n}}\left( t \right)} \right) = H\left( {{s_{i,n}}\left( t \right) - \frac{1}{2}S} \right),\\
  &{C_2}\left( {t,{s_{i,n}}\left( t \right),{s_{ - i,n}}\left( t \right)} \right)=\\
  &\ \ \ \ \left( {1 - H\left( {{s_{i,n}}\left( t \right) - \frac{1}{2}S} \right)} \right)H\left( {{s_{ - i,n}}\left( t \right) - S} \right),\\
     &{C_3}\left( {t,{s_{i,n}}\left( t \right),{s_{ - i,n}}\left( t \right)} \right) =\\
     &\  \ \ \ \left( {1 - H\left( {{s_{i,n}}\left( t \right) - \frac{1}{2}S} \right)} \right)\left( {1 - H\left( {{s_{ - i,n}}\left( t \right) - S} \right)} \right),
  \end{split}
   \end{equation*}
   where $H\left( x \right) = \left\{ \begin{array}{l}
0,x < 0\\
1,x \ge 0
\end{array} \right.$ is the heaviside step function.}

 {The first term in (\ref{Delay}) denotes the delay when the user is served by its default F-AP ${i}$, which corresponds to case $1$ and $3$. Hence, it is the product of the occurrence condition and the delay under the condition, which refers to the wireless transmission delay between F-AP $i$ and user $k$ and the fronthaul delay induced by retrieving  the uncached segment of ${f_n}$ from the cloud server. The second term in (\ref{Delay}) denotes the delay when the user is served by some adjacent F-AP in case 2, and the corresponding delay refers to the wireless transmission delay brought by the other serving F-AP. Note that ${{{s}_{-i,n}}\left( t \right)}$ and ${{{R}_{ - i,k}}\left( t \right)}$ can be approximated by the average states of all the other F-APs denoted by ${{{\bar s}_{-i,n}}\left( t \right)}$ and ${{{\bar h}_{-i,k}}\left( t \right)}$, which will be presented in Section III-C.}

\subsubsection{Fronthaul Traffic Load}
 The traffic load on the fronthual link is induced by both content caching in the caching phase and content retrieving in the delivery phase. {Let ${O_{i,n}}\left( t \right)$ denote the fronthaul load of F-AP ${i}$ with respect to ${f_n}$ at time $t$, and it can be expressed in (\ref{Load}) at the top of this page. The first term in (\ref{Load}) characterizes the instantaneous fronthaul load in the content caching process with \({{\eta _1}}\) and \({{\eta _2}}\) as parameters, and it is a simple quadratic function that is monotonous increasing with the caching rate ${c_{i,n}}\left( t \right)$. The second term in (\ref{Load}) is the fronthaul load generated by content retrieving when users that request for $f_n$ are served by F-AP $i$, i.e., case 1 and case 3. In these two cases, F-AP ${i}$ has to repeatedly retrieve the uncached file segment through the fronthual link for each request; hence, the fronthaul load is the product of the request number, the size of the file segment to be retrieved and the occurrence condition, while $\eta $ is a numerical coefficient.}

Given (\ref{Delay}) and (\ref{Load}), the total instantaneous cost \({J_{i,n}}\left( t \right)\) of F-AP ${i}$ can be represented by the weighted summation of the request service delay and the fronthaul traffic load, {and it can be expressed as follows:}
\setcounter{equation}{5}
 \begin{equation}\label{Cost}
   {J_{i,n}}\left( t \right) = {{{\omega}_1}D_{i,n}}\left( t \right) + {{{\omega}_2}O_{i,n}}\left( t \right),
\end{equation}
where $\omega_1$ and $\omega_2$ are the coefficients to bring the two corresponding terms to the same scale. Define \({{{\bm{J}}_i}\left( t \right)} = {\left[ {{J_{i,1}\left( t \right)},{J_{i,2}\left( t \right)}, \cdots ,{J_{i,N}\left( t \right)}} \right]^{\rm{T}}}\). Then,  it denotes the cost vector of F-AP $i$ at time $t$ for all files.

\subsection{Stochastic Differential Game Formulation}
Consider the caching placement optimization in a finite time horizon $\left[ {0,T} \right]$. As can be observed from (\ref{Delay}) and (\ref{Load}), the cost function of each F-AP is determined by states of all the F-APs; hence the states and policies of F-APs have direct impacts on each other. Therefore, the caching optimization problem of F-APs can be characterized by a stochastic differential game that captures the interactive relationship among the F-APs.

The stochastic differential game is defined by a 4-tuple $\left( {\mathcal{\boldsymbol{I}}}, {{\left\{ {{\mathcal{\boldsymbol{X}}_i}} \right\}}_{i \in \mathcal{\boldsymbol{I}}}}, {{\left\{ {{\mathcal{\boldsymbol{C}}_i}} \right\}}_{i \in \mathcal{\boldsymbol{I}}}}, {{\left\{ {{\mathcal{\boldsymbol{J}}_i}} \right\}}_{i \in \mathcal{\boldsymbol{I}}}} \right)$.  ${\mathcal{\boldsymbol{I}}}$ is the set of players, which in this case are F-APs; ${\mathcal{\boldsymbol{X}}_i}$ denotes the state set of player ${i}$, which is the space of possible state vector ${\bm{x}_i}$; ${\mathcal{\boldsymbol{C}}_i}$ denotes the action set of player ${i}$, which represents the set of possible caching policy vector ${{\bm{c}}_i}$; ${\mathcal{\boldsymbol{J}}_i}$ denotes the cost of player ${i}$ for all files over the time horizon with $\mathcal{\boldsymbol{J}}_i = \mathbb{E}\left[ {\int_0^T {{{\bm{J}}_i}\left( t \right)dt} } \right]$.

In the SDG established above, the caching optimization problem of player $i$ can be expressed as follows:
\begin{equation}\label{3-4}
  \mathop {\min }\limits_{{{\bm{c}}_i}\left( {0 \to T} \right)} {\mathcal{\boldsymbol{J}}_i}\left( {{{\bm{c}}_i},{{\bm{x}}_i},{{\bm{x}}_{ - i}}} \right),
\end{equation}
where ${{{\bm{c}}_i}\left( {0 \to T} \right)}$ denotes the caching policy over the time horizon, and ${{\bm{x}}_{ - i}} = \left\{ {{\bm{x}_j}, j \ne i, j \in \mathcal{\boldsymbol{I}}} \right\}$ denotes the states of the other $I-1$ players.

According to the theory of dynamic programming, the overall optimum of a problem for the entire time horizon $\left[ 0,T \right]$ can be obtained by sequentially constructing the optimal policies of its subproblems over time period $\left[ t,T \right]$ in a time-reversed manner \cite{Bertsekas}. Correspondingly, we define the value function ${v_{i,n}}\left( t \right)$ of player $i$ w.r.t. ${f_n}$ at time $t$ as the minimal cost over time period $\left[ t,T \right]$ as follows:
\begin{equation}\label{3-5}
  {v_{i,n}}\left( {t} \right) = \mathop {\min }\limits_{{c_{i,n}\left( {t \to T} \right)}} \mathbb{E}\left[ {\int_t^T {{J_{i,n}}\left( r \right){\rm d}r} } \right].
\end{equation}

In the SDG, the set of caching policies \(\left\{ {{\bm{c}}_i^*\left( {0 \to T} \right)}, i \in \mathcal{\boldsymbol{I}} \right\}\) that minimizes the cost functions constitutes a feedback Nash equilibrium (NE), and it is guaranteed if there exist value functions that satisfy the following backward Hamilton-Jacobi-Bellman (HJB) partial differential equations (PDEs) \cite{Yeung}:
\begin{equation}\label{3-7}
  {\partial _t}{v_{i,n}}\left( t \right) + \mathop {\min }\limits_{{c_{_{i,n}}}\left( t \right)} \left[ {{L}{v_{i,n}}\left( t \right) + {J_{i,n}}}\left( t \right) \right] = 0, \forall i \in \mathcal{\boldsymbol{I}},{f_n} \in \mathcal{\boldsymbol{F}},
\end{equation}
where \({{L}{v_{i,n}}\left( t \right)}\) denotes the differential operator of the value function.

To solve the coupled $I$ HJB
equations for each file, however, requires each F-AP to acquire the state information of all the other F-APs, which needs high computational complexity and induces enormous extra communicating overhead. Therefore, we propose to solve this challenging problem by introducing mean field game approximation.
\subsection{The Proposed Mean Field Edge Caching Scheme}

The intractability of the solution to the original SDG due to the coupled PDEs can be addressed by an approximation of the original game as a mean field game, in which individual states of all the players can be captured by a single statistical distribution \cite{MFG}. Therefore, the number of the coupled PDEs can be significantly reduced and the players are able to make decisions in a distributed manner.

The establishment of a mean field game is based on four conditions \cite{Gueant}: (1) rationality of players; (2) continuum of players; (3) exchangeability among states of players; (4) interactions among players of the mean field type.
The first condition can be guaranteed since each F-AP makes logical decisions. The second condition can be assured according to the ultra-dense deployment of F-APs. The third condition holds since the contribution of individuals become infinitesimal among the mass of players and the interchange among F-APs indices does not alter the outcome of the game. The fourth condition is assured since each F-AP interacts with the mean field instead of actual individuals in the game.

Given the properties presented above, the original SDG can be approximated as an MFG, in which each player is indistinguishable among the mass of players. Therefore, we can consider a generic player by dropping the F-AP index. The interactions between the generic player with the others in the MFG are sufficiently characterized by its own state ${\bm{x}}$ and the statistical distribution of the mass state ${\bm{m}}$. ${\bm{x}} = \left( {{\bm{h}},{\bm{s}}} \right)$ denotes the state of the generic player. ${\bm{m}}$ represents the mean field distribution, and its element w.r.t. file ${f_n}$ can be expressed as follows:
\begin{equation}\label{4-1}
  {m _n}(t) = \mathop {\lim }\limits_{I \to \infty } \frac{1}{I}\sum\limits_{i = 1}^I {{\delta _{{{\bm{x}}_{i,n}}(t)}}},
\end{equation}
where ${{\delta _{{{{x}}_{i,n}}(t)}}}$ is the Dirac measure. Note that the mean field distribution is the probability distribution of the states of all the players in the game. The evolution of ${m _n}(t)$ corresponds to the Fokker-Planck-Kolmogorov (FPK) equation as follows \cite{Lasry}:
\begin{equation}\label{4-2}
  \begin{split}
  {\partial _t}{m_n}\left( t \right) & = \frac{1}{2}{\sigma^2 _h}\partial _{hh}^2{m_n}\left( t \right)- \frac{1}{2}\alpha({\mu _h} - h\left( t \right)){\partial _h}{m_n}\left( t \right)\\
 & - S\left( {{c_n}\left( t \right) - {e^{a - 1}}{a^{{q_n}\left( t \right)}}} \right){\partial _s}{m_n}\left( t \right).
\end{split}
\end{equation}
Given the mean field distribution, the average states of other players in the mean field can be approximated as the functions of ${m_n}(t)$ as follows:
\begin{equation}\label{4-7}
  \left\{ \begin{array}{l}
{{\bar h}_{ - ,k}}\left( t \right) = \int {{h_k}\left( t \right){m_n}(t){\rm d}{h_k}}, \\
{{\bar s}_{ - ,n}}\left( t \right) = \int {{s_n}\left( t \right){m_n}(t){\rm d}{s_n}}.
\end{array} \right.
\end{equation}

With the establishment of the mean field, each player can obtain its policy based solely on the local information, including its local state and the mean field distribution. Correspondingly, the objective of each player can be reformulated as a generic problem as follows:
\begin{equation}\label{4-3}
  \mathop {\min }\limits_{\bm{c}\left( {0 \to T} \right)} \mathcal{\boldsymbol{J}}\left( {{\bm{c}},{\bm{x}},{\bm{m}}} \right).
\end{equation}

The solution of (\ref{4-3}) can be obtained by solving the following HJB equation:
\begin{equation}\label{4-4}
  {\partial _t}{v_n}\left( t \right) + \mathop {\min }\limits_{{c_{_n}}\left( t \right)} \left[ {{L}{v_n}\left( t \right) + {J_n}\left( t \right)} \right] = 0,
\end{equation}
where the differential operator ${L}{v_n}\left( t \right)$ can be expressed as follows:
\begin{equation}
\begin{split}
{L}{v_n}\left( t \right) & = \frac{1}{2}{\sigma^2 _h}\partial _{hh}^2{v_n}\left( t \right) + \frac{1 }{2}\alpha ({\mu _h} - h\left( t \right)){\partial _h}{v_n}\left( t \right)\\
 &+ S\left( {{c_n}\left( t \right) - {e^{a - 1}}{a^{{q_n}\left( t \right)}}} \right){\partial _s}{v_n}\left( t \right).
 \end{split}
\end{equation}

Here we define the Hamiltonian function of the optimization problem as \(G\left( {t,{c_n},{\bm{x}}} \right) = {L}{v_n}\left( t \right) + {J_n}\left( t \right)\). As can be observed from (\ref{4-4}), the optimal caching policy $c_n^*\left( t \right)$ can be obtained by minimizing the Hamiltonian and concretely derived as follows:
\begin{equation}\label{4-6}
 c_n^*\left( t \right) =  - \frac{1}{{{\eta _2}}}\left( {\frac{S}{{{\omega _2}}}{\partial _s}{v_n} + {\eta _1}} \right).
\end{equation}

Clearly, the solution of the HJB equation affects the FPK equation through the caching policy $c_n\left( t \right)$, while the solution of the FPK equation determines the HJB equation through the cost function $J_n(t)$. The two coupled PDEs comprehensively form the structure of the mean field: the backward HJB equation models the induction process of optimization of each individual, while the forward FPK equation models the evolution of the mean field as a whole.
The solutions, i.e., ${m_n}\left( t \right)$ and ${v_n}\left( t \right)$, of the FPK-HJB equations in (\ref{4-2}) and (\ref{4-4}) constitute the NE of the MFG. The existence and uniqueness of the NE are guaranteed by the smoothness of the Hamiltonian \cite{Lasry}, which can be easily verified.

Note that the primary $I$ coupled equations have been reduced to two simple equations that can sufficiently characterize the game. Each F-AP can iteratively solve the above two equations till convergence to obtain the optimal caching policy with respect to each file. In the case that the overall cache state of F-AP $i$ exceeds the space limit $C$, the files with the minimal obtained $v_n(t)$ can be chosen to meet the constraint. Correspondingly, the optimization can be executed in a distributed manner without observing the specific states of other individual F-APs.

\section{Simulation Results}
The proposed distributed edge caching scheme is verified through simulations, in which a circular area with the radius of 600 m is considered. 
The coupled HJB-FPK equations are iteratively solved to the fixed point by utilizing the PDEPE toolbox of MATLAB. Due to the close distance between users and F-APs in ultra-dense F-RAN, a static channel model is assumed in the simulations, thus the dynamic states of the F-APs can be characterized simply by the cache states \cite{Distr4}. Moreover, the time horizon $T$ is assumed to be normalized. Other parameters are set as follows: $P_i = \text{1 W}$, $\sigma ^2 = - 77\text{ dBm}$, $W = \text{10 MHz}$, $K = 100$, $S = 100\text{ MB}$, ${\eta _1} = 0.0001, {\eta _2} = 0.05$, $\eta = 0.3$, $\omega_1 = 100$, $\omega_2 = 10^{-6}$. 

\begin{figure}[t]
 \centering
\includegraphics[width=\linewidth]{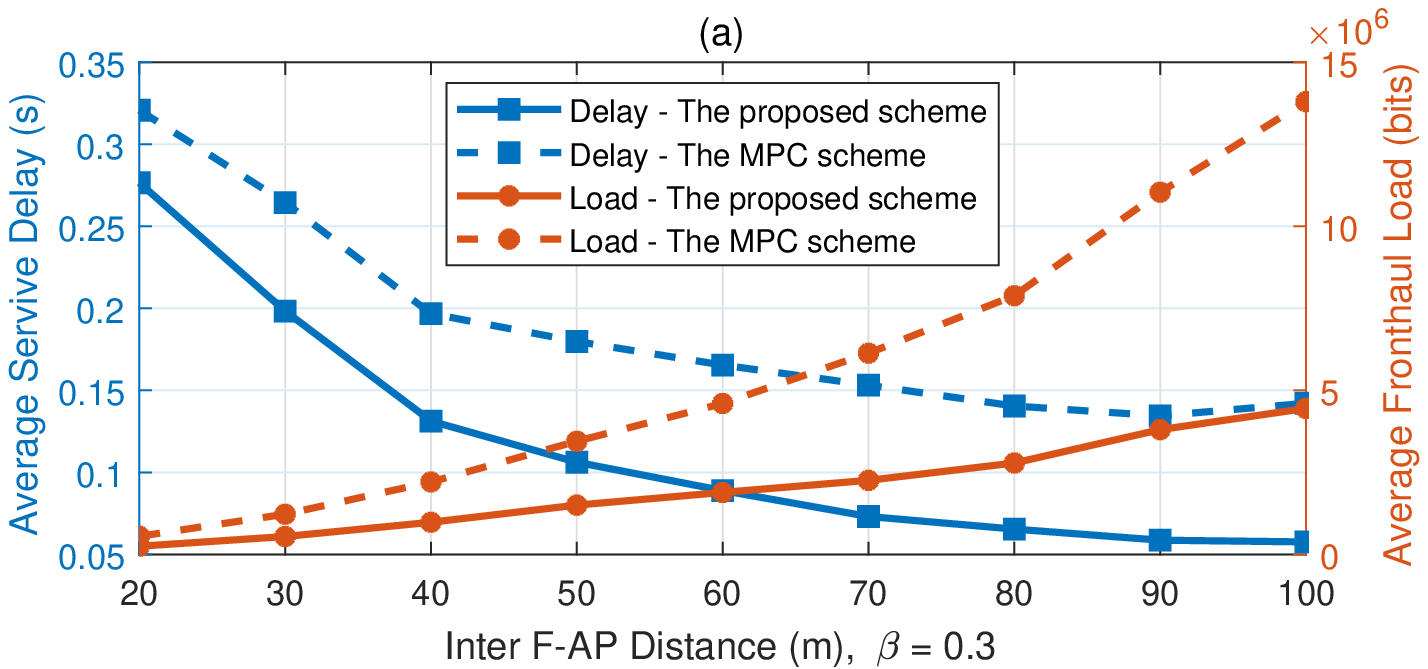}\\
\includegraphics[width=\linewidth]{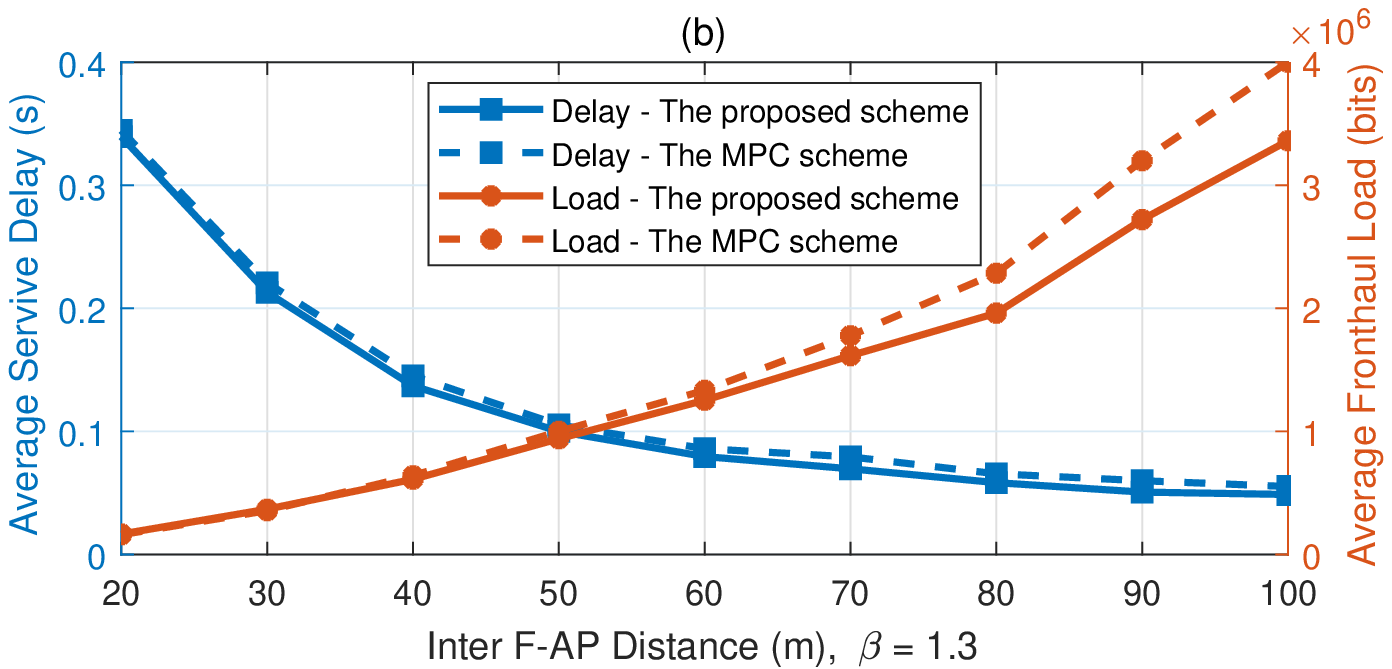}
\caption{Average request service delay and fronthaul traffic load versus inter F-AP distance for the proposed and MPC schemes under static user requests with \(\beta  = 0.3 \text{ and } \beta  = 1.3\).}
\label{Static}
\end{figure}

Assume that the probability of a user requesting file ${f_n}$ follows Zipf distribution, i.e., ${p_n} = \frac{{{n^{ - \beta }}}}{{\sum\nolimits_{n = 1}^N {{n^{ - \beta }}} }},$
 where the parameter $\beta$ depicts the steepness of the distribution with a positive value and a larger value of $\beta$ implicates a more skewed distribution of content popularity. In Fig. \ref{Static}, we show the average request service delay and fronthaul traffic load versus inter F-AP distance (IFD) of the proposed scheme and the most popular caching (MPC) scheme under static user requests with $\beta=0.3$ and $\beta=1.3$, respectively. As shown, the average request service delay decreases with IFD, while the average fronthaul traffic load of F-APs increases with IFD. The reason for this behavior is that the F-RAN becomes sparse and the inter F-AP interference decreases as IFD increases, which result in increased wireless transmission rate and requests number of each F-AP. Moreover, Fig. 1(a) shows an obvious performance gap between  the proposed scheme and the MPC scheme, while the performance gain is relatively small in Fig. 1(b). This is a result of the fact that the MPC scheme performs well only when the preference of users can be depicted by the fixed Zipf distribution. On the contrary, the proposed scheme shows satisfactory performance with different $\beta$, because it is able to capture the real-time request status even when users show less obvious preference for certain contents.

\begin{figure}[t]
\centering 
\includegraphics[width=\linewidth]{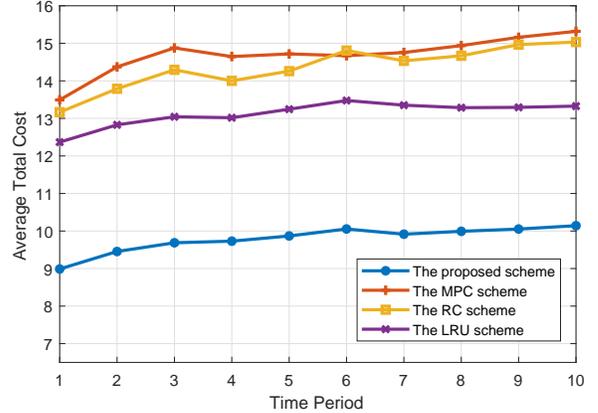}
\caption{Average total cost versus time period for the proposed and baseline schemes under time-variant user requests.}
\label{TimeVariant_Total}
\vspace{-0.5cm}
\end{figure}

In Fig. \ref{TimeVariant_Total}, we show the average total cost of the proposed and baseline schemes under time-variant user requests, where the content popularity distribution is randomized in consecutive time periods. We choose the MPC, random caching (RC), and least recently used (LRU) \cite{LRU} schemes as three baselines. It can be observed that the proposed scheme shows a stable performance gain over the baseline schemes. Specifically, the average total cost is reduced by $33\%$ and $25\%$ compared with the MPC and LRU schemes, respectively. 
The reason is that the MPC and RC schemes employ static caching policies without considering time-variant user requests, and the LRU scheme employs dynamic caching policies based only on local cache state. On the contrary, the proposed scheme optimizes the caching policies according to real-time requests and takes into account both local state and mean field distribution. Therefore, the proposed scheme can adapt to time-variant user requests to minimize the average total cost.

\section{Conclusions}
In this paper, we have proposed a dynamic distributed edge caching scheme in ultra-dense F-RAN by considering time-variant user requests. By utilizing MFG theory, the proposed scheme enables each F-AP to determine its dynamic caching policy based on local information to minimize both the request service delay and fronthaul traffic load. The simulation results have shown that the proposed scheme provides considerable performance improvement over the baseline schemes. 

\section*{Acknowledgments}

This work was supported in part by
the Research Fund of the State Key Laboratory of
Integrated Services Networks (Xidian University) under grant ISN19-10,
the Research Fund of the Key Laboratory of Wireless Sensor Network $\&$ Communication (Shanghai Institute of Microsystem and Information Technology, Chinese Academy of Sciences) under grant 2017002,
the Hong Kong, Macao and Taiwan Science $\&$ Technology Cooperation Program of China under grant 2014DFT10290,
the Ericsson and SEU Cooperation Project under grant 8504000335,
the National Basic Research Program of China
(973 Program) under grant 2012CB316004,
and the U.K. Engineering and Physical Sciences Research Council under Grant EP/K040685/2.

\bibliographystyle{IEEEtran}
\bibliography{Manuscript-CachingMFG-VTC2018FALL}

\end{document}